\documentclass[12pt,onecolumn]{IEEEtran}

\usepackage[colorlinks=false]{hyperref}
\usepackage[dvips]{graphicx}
\usepackage[usenames,dvipsnames]{color}
\usepackage{epsfig}
\usepackage{rotating}
\usepackage{amssymb}
\usepackage[boxed]{algorithm}
\usepackage[latin1]{inputenc}
\usepackage{verbatim}

\begin{document}

\newcommand{\beg}{\begin{equation}}
\newcommand{\eeg}{\end{equation}}
\newcommand{\br}{\begin{array}}
\newcommand{\er}{\end{array}}
\newcommand{\bea}{\begin{equation} \begin{array}{c}}
\newcommand{\eea}{\end{array} \end{equation}}
\newcommand{\un}{\underline}
\newcommand{\cb}{\begin{center}}
\newcommand{\ce}{\end{center}}
\newcommand{\llg}{\left\langle}
\newcommand{\rrg}{\right\rangle}
\newcommand{\thh}{$^{\mbox{th}}$}
\newcommand{\al}{\alpha}
\newcommand{\bite}{\begin{itemize}}
\newcommand{\eite}{\end{itemize}}
\newcommand{\m}{\mbox}
\newcommand{\wh}{\hspace{3mm} \mbox{ where }}
\newcommand{\lcm}{\mbox{lcm}}
\newcommand{\di}{\mbox{ div }}
\newcommand{\T}{\mbox{ TRUE }}
\newcommand{\SP}{\mbox{ SPAN}}
\newcommand{\GF}{\mbox{ GF}}
\newcommand{\gf}{\mbox{ {\tiny{GF}}}}
\newcommand{\GR}{\mbox{ GR}}
\newcommand{\Di}{\mbox{ DIV}}
\newcommand{\mt}{\hspace{10mm} \mbox{ }}
\newcommand{\mf}{\hspace{5mm} \mbox{ }}
\newcommand{\mz}{\hspace{2mm} \mbox{ }}
\newcommand{\mv}{\hspace{1mm} \mbox{ }}
\newcommand{\ga}{\gamma}
\newcommand{\xl}{\begin{tiny} \begin{array}{c} < \\ \simeq \end{array} \end{tiny}}
\newcommand{\xm}{\begin{tiny} \begin{array}{c} > \\ \simeq \end{array} \end{tiny}}
\newcommand{\OR}{\mbox{ ord}}
\newcommand{\dg}{\mbox{ deg}}
\newtheorem{thm}{Theorem}
\newtheorem{pro}{Proposition}
\newtheorem{lem}{Lemma}
\newtheorem{cj}{Conjecture}
\newtheorem{cor}{Corollary}
\newtheorem{df}{Definition}
\newtheorem{imp}{Implication}
\newcommand{\mo}{\mbox{ mod }}
\newcommand{\Tr}{\mbox{ Tr}}
\newcommand{\mn}{\mbox{ {\tiny{min}}}}
\newcommand{\mx}{\mbox{ {\tiny{max}}}}
\newcommand{\erf}{\mbox{ erf}}
\newcommand{\erfc}{\mbox{ erfc}}
\newcommand{\SNR}{\mbox{ SNR}}
\newcommand{\BER}{\mbox{ BER}}
\newcommand{\hl}{\\ \hline}
\newcommand{\bd}[1]{\textbf{#1}}

\title{Generalised Bent Criteria for Boolean Functions (II)}
\author{Constanza Riera\thanks{C. Riera is with the Depto. de \'{A}lgebra,
Facultad de Matem\'{a}ticas, Universidad Complutense de Madrid,
Avda. Complutense s/n, 28040 Madrid, Spain. E-mail: \texttt{criera@mat.ucm.es}. Supported by the Spanish Government Grant AP2000-1365.},
George Petrides\thanks{G. Petrides is with the School of Mathematics,
University of Manchester, P.O. Box 88, Sackville Street, Manchester, M60 1QD, UK. E-mail: \texttt{george.petrides@manchester.ac.uk}.
Supported by the Marie Curie Scholarship.} and 
Matthew G. Parker\thanks{M. G. Parker is with the
  Selmer Centre, Inst. for Informatikk, H{\o}yteknologisenteret i Bergen,
  University of Bergen, Bergen 5020, Norway. E-mail: \texttt{matthew@ii.uib.no}.
  Web: \texttt{http://www.ii.uib.no/\~{}matthew/}}}

\date{\today}
\maketitle


\begin{abstract}
In the first part of this paper \cite{RP:BC}, some results
on how to compute the flat spectra of Boolean
constructions w.r.t. the transforms $\{I,H\}^n$, $\{H,N\}^n$ and $\{I,H,N\}^n$
were presented,
and the relevance of Local Complementation to the quadratic case was indicated.
In this second part, the results are applied to develop recursive formulae for
the numbers of flat spectra of some structural quadratics.
Observations are made as to the generalised
Bent properties of boolean functions of algebraic degree greater than two,
and the number of flat spectra w.r.t. $\{I,H,N\}^n$ are computed for
some of them.
\end{abstract}

\section{Introduction}
In this work, we apply the techniques that we presented in Part I \cite{RP:BC}, to prove that, for certain recursive quadratic boolean
constructions,
one can establish simple recursive relationships for the number of flat spectra
w.r.t. the $\{I,H,N\}^n$ transform set. For those boolean constructions, we prove simple recursions for the number of flat spectra
w.r.t. the $\{I,H,N\}^n$ transform set or subsets
thereof. We also observe that optimal {\em{Quantum Error-correcting Codes}} (QECCs), interpreted as quadratic
boolean functions, appear to maximise the number of flat spectra w.r.t.
$\{I,H,N\}^n$. Very loosely, for boolean functions of fixed degree,
the more flat (or near-flat) spectra
w.r.t. $\{I,H,N\}^n$ we obtain for the function, the stronger
it is cryptographically, and the more {\em{entangled}} it is when
interpreted as a quantum state \cite{Par:QE,Hein:GrEnt} - these measures of
cryptographic strength and/or entanglement are only partial.

Before presenting our results, we will recapitulate the sections of Part I which are helpful to the understanding of Part II.

Let $H = \frac{1}{\sqrt{2}}\begin{tiny} \left ( \begin{array}{rr}
1 & 1 \\
1 & -1
\end{array} \right ) \end{tiny}$ be the Walsh-Hadamard kernel, 
$N = \frac{1}{\sqrt{2}}\begin{tiny} \left ( \begin{array}{rr}
1 & i \\
1 & -i
\end{array} \right ) \end{tiny}$, where $i^2 = -1$, the Negahadamard kernel, and $I$ the $2\times2$  identity matrix. We say that a Boolean function \linebreak
$p({\bf{x}}):\mbox{GF}(2)^n\rightarrow \mbox{GF}(2)$ is {\em Bent} \cite{Rot:Bent} if $P=2^{-n/2}(\bigotimes_{i=0}^{n-1} H)
(-1)^{p({\bf x})}$ has a {\em{flat}} spectrum, or, in other words, if $P=(P_{\bf{k}}) \in {\mathbb{C}}^{2^n}$ is such that
$|P_{\bf{k}}| = 1\ \forall \ {\bf{k}}\in \mbox{GF}(2)^n$. Bent boolean functions are desirable cryptographic primitives as they optimise
resistance to linear cryptanalysis.
If the function is quadratic, we can associate to it a simple non-directed
graph, and in this case a flat spectrum is obtained iff $\Gamma$, the adjacency matrix of the graph,  has maximum rank mod 2.
In Part I, we generalised this concept, considering not only the Walsh-Hadamard transform $\bigotimes_{i=0}^{n-1} H$,
but the complete
set of unitary transforms $$\{I,H,N\}^n=\bigotimes_{j \in {\bf{R_I}}} I_j \bigotimes_{j \in {\bf{R_H}}} H_j
\bigotimes_{j \in {\bf{R_N}}} N_j \enspace,$$ where the sets ${\bf{R_I}}, {\bf{R_H}}$ and ${\bf{R_N}}$ form a partition of the set of vertices $\{0,\ldots,n-1\}$. With this generalised criterion, we study the number of flat spectra of a function w.r.t. $\{I,H,N\}^n$. We also consider the number of flat spectra w.r.t. some subsets of $\{I,H,N\}^n$, namely $\{H,N\}^n$ (when ${\bf{R_I}}=\emptyset$) and $\{I,H\}^n$ (when ${\bf{R_N}}=\emptyset$).  Note that the Walsh-Hadamard transform, $\{H\}^n$, is the intersection of these subsets. We prove that a function will have a flat spectrum w.r.t. a transform in $\{I,H,N\}^n$ iff a certain modification of its adjacency matrix, concretely the matrix resultant of the following actions, has maximum rank mod 2:
\vspace{3mm}
\begin{itemize}
\item for $i\in{\bf{R_I}}$, we erase the $i^{th}$ row and column 
\item for $i\in{\bf{R_N}}$, we subsitute 0 for 1 in position $[i,i]$ 
\item for $i\in{\bf{R_H}}$, we leave the $i^{th}$ row and column unchanged.
\end{itemize}
\vspace{3mm}

In sections \ref{HN}, \ref{IH} and \ref{IHN}, we compute, by means of this modified matrix and using the result exposed above,
the number of flat spectra for some Boolean functions w.r.t $\{H,N\}^n$, $\{I,H\}^n$ and $\{I,H,N\}^n$ respectively.
It is desirable to identify boolean functions which maximise
the number of flat spectra w.r.t. $\{I,H,N\}^n$, as this is an
indicator of high {\em{entanglement}} for the
corresponding pure multipartite quantum states which are represented by the
boolean functions \cite{Par:QE,Hein:GrEnt} (see Part I \cite{RP:BC}).
We will see that the quadratic line and clique functions appear to maximise the number of flat spectra w.r.t.
$\{H,N\}^n$ and $\{I,H\}^n$, respectively, and that the quadratic functions representing high-distance QECCs appear
to maximise the number of flat spectra w.r.t. $\{I,H,N\}^n$. Recent graphical descriptions for these optimal
QECCs \cite{DanPAR} suggest that {\em{nested-clique}} structures may maximise the number of flat spectra w.r.t. $\{I,H,N\}^n$.
As an initial step towards the analysis of such functions we provide recursive formulae for the number
of flat spectra for the 'clique-line-clique' structure.

Some recent papers \cite{Arr:Int,Aig:Int,Arr:Int1,Arr:Int2}
have proposed {\em{interlace polynomials}} to describe interlace/circle graphs.
In particular, polynomials 
$q(x)$ and $Q(x)$ are defined, and proved to be certain {\em{Martin polynomials}},
as proposed by Bouchet \cite{Bou:Mart}. It can be shown that $q(x)$ and $Q(x)$
summarise certain aspects of the spectra of a graph w.r.t.
$\{I,H\}^n$ and $\{I,H,N\}^n$, respectively. In particular, $q(1)$ and $Q(2)$
evaluate the number of flat spectra
w.r.t. $\{I,H\}^n$ and $\{I,H,N\}^n$, respectively. In this paper we will
point out links with this work but defer a thorough investigation to a future
paper.

Section \ref{conc} contains a few concluding remarks;
finally, we give tables summarising our results in the appendix (section \ref{tables}).

\section{On the Number of Flat Spectra of Quadratic Boolean Functions with respect to $\{H,N\}^n$} \label{HN}

We wish to construct boolean functions that have flat spectra w.r.t. the largest possible subset of
$\{H,N\}^n$ transforms. The multivariate complementary set constructions
of \cite{Par:LowPAR} provide candidate functions. The simplest and strongest of these is the
line function (or path graph) \cite{Rud:RS,Gol:Comp,Dav:PF}.

\subsection{Line}
The {\em{line function}}, $p_l({\bf{x}})$ is defined as
\beg p_l({\bf{x}}) = \sum_{j=0}^{n-2} x_jx_{j+1} + {\bf{c \cdot x}} + d \enspace,
\label{RS} \eeg
where ${\bf{x}},\ {\bf{c}} \in \mbox{GF}(2)^n$, ${\bf{x}}=(x_0,\ldots,x_{n-1})$, and $d \in \mbox{GF}(2)$. Its number of flat spectra with respect to
$\{H,N\}^n$ is as follows:

\begin{lem}\label{lem:lineHN}
$ K_n=\#\mbox{ flat spectra}(p_l({\bf{x}})) \mbox{ w.r.t. } \{H,N\}^n =
2^n-K_{n-1},\mbox{ with }K_1=1$; in closed form, $$K_n=\frac{1}{3}\left(2^{n+1}+(-1)^n\right)\enspace.$$\end{lem} 
\begin{proof}
The generic modified matrix of the line for $\{H,N\}^n$ is as follows:

\begin{small}
$$\Gamma_{\bf{v}}=\left(\begin{array}{ccccccc}
v_0 & 1 & 0 & 0 & \ldots & 0 & 0\\
1 & v_1 & 1 & 0 & \ldots & 0 & 0\\
0 & 1 & v_2 & 1 & \ldots & 0 & 0\\
\vdots & \vdots & \vdots & \vdots & \ddots & \vdots & \vdots\\
0 & 0 & 0 & 0 & \ldots & 1 & v_{n-1}\end{array}\right)\enspace,$$
\end{small}
where ${\bf{v}}=(v_0,\ldots,v_{n-1})\in \mbox{GF}(2)^n$.

Computing the determinant, we get the recursion formula $$D_n=v_{0}D_{n-1}+D_{n-2} \mbox{ mod }2\enspace,$$
where $D_{n-j}$ is the determinant of the generic modified matrix of the line in the variables $x_j,\ldots,x_{n-1}$.
The spectra will be flat iff $D_n=1$. In order to get this, we consider the following cases:
\begin{enumerate}
\item $D_{n-1}=0, D_{n-2}=1$. In this case, $v_0$ can be 0 or 1.
\item $D_{n-1}=1, D_{n-2}=1$. In this case, $v_0=0$.
\item $D_{n-1}=1, D_{n-2}=0$. In this case, $v_0=1$.
\end{enumerate}

We then have $K_n=2N1+N2+N3$, where $Ni$ is the number of times the $i^{th}$
case is true. Note that
$\{v_1,\ldots,v_{n-1}|D_{n-1}=D_{n-2}=1\}\cup \{v_1,\ldots,v_{n-1}|D_{n-1}=
1, D_{n-2}=0\}= \{v_1,\ldots,v_{n-1}|D_{n-1}=1\}$,
and therefore $N2+N3=K_{n-1}$.

We see now that $\{v_1,\ldots,v_{n-1}|D_{n-1}=0, D_{n-2}=1\}=
\{v_1,\ldots,v_{n-1}| D_{n-1}=0\}$, and so \linebreak 
$N1=2^{n-1}-K_{n-1}$. 
Suppose $D_{n-1}=D_{n-2}=0$. As $ D_{n-1}=v_{1}D_{n-2}+D_{n-3}$, this implies $D_{n-3}=0$. By the same argument, we must have $D_i=0,\ 1 \leq i \leq n-1$. However, if $D_1= v_{n-1}=0$ then $D_2=v_{n-2}v_{n-1}+1=1$, and this leads to a contradiction.

Finally, $K_n=2(2^{n-1}-K_{n-1})+K_{n-1}=2^n-K_{n-1}$.
Expanding this recurrence relation, and using $N_0=1$, we get
$K_n=\displaystyle \sum_{k=0}^n(-1)^{n+k}2^k=\frac{1}{3}\left(2^{n+1}+(-1)^n\right)$\enspace.
\end{proof}

\subsection{Clique}
We define the {\em{clique function}} (that is, the {\em{complete graph}}) as,
\beg p_c({\bf{x}}) = \sum_{0\leq i < j\leq n-1} x_ix_j \enspace,\label{clique} \eeg 
where ${\bf{x}}=(x_0,\ldots,x_{n-1})\in \mbox{GF}(2)^n$. For this function, the number of flat spectra with respect to
$\{H,N\}^n$ is given as follows:

\begin{lem} \label{lem:cliqueHN} $ K_n=\#\mbox{ flat spectra}(p_c({\bf{x}}))
\mbox{ w.r.t. } \{H,N\}^n = K_{n-1}+1+(-1)^n$; in closed form,
$$ K_n=n+\frac{1+(-1)^n}{2}\enspace.$$ \end{lem}
\begin{proof}
The generic modified adjacency matrix of the clique is as follows:

\begin{small}
$$\Gamma_{\bf{v}}=\left(\begin{array}{cccccc}
v_0 & 1 & 1 & 1 & \ldots & 1\\
1 & v_1 & 1 &  1 & \ldots & 1\\
1 & 1 & v_2 & 1 & \ldots & 1\\
\vdots & \vdots & \vdots & \vdots & \ddots & \vdots\\
1 & 1 & 1 & 1 & \ldots & v_{n-1}\end{array}\right)\enspace.$$
\end{small}

Applying $N$ to the bipolar vector of the clique function, $(-1)^{p_c({\bf{x}})}$, in the position $i$ is equivalent to making $v_i=1$.
If two or more of the $v_i$'s are 1, then the matrix will not have full rank, so $|{\bf{R_N}}|\leq 1$. 

First, suppose $|{\bf{R_N}}|=1$. Computing the determinant, we get $D=\mbox{det}(\Gamma_{\bf{v}})=\mbox{det}(\Gamma)+m$,
where $m$ is the minor corresponding to $v_i$. Obviously, $m$ is the determinant of the adjacency matrix of a clique
in $n-1$ variables. It is easy to show that the clique in $n$ variables is bent iff $n$ is even.
So, if $n$ is even, we have $\mbox{det}(\Gamma)=1,\ m=0$, and so $D=1$. On the other hand, if $n$ is odd,
we have $\mbox{det}(\Gamma)=0,\ m=1$, and so $D=1$. This means that for every position in which we choose to apply $N$,
we have a flat spectrum, and therefore we get $n$ flat spectra for this case.

Now, suppose $|{\bf{R_N}}|=0$. Since the clique is bent in an even number of variables, we have flat spectra iff $n$ is even.

From the preceding argument, we see that $ K_n=n+\frac{1+(-1)^n}{2}  $.
The recurrence formula follows trivially.
\end{proof}

\subsection{Clique-Line-Clique}
By combining the clique and line graphs in certain ways we can get an improvement in the
number of flat spectra of a clique in the same number of variables,
though we are still far from the number of flat spectra of a line in the same number of variables. 

Specifically, if we define the
$n$ {\em clique-line-$m$ clique} as
\beg p_{n,m}({\bf{x}})=\displaystyle \sum_{_{0\leq i < j\leq n-1}} x_ix_j+x_{n-1}x_{n}+\sum_{_{n\leq i < j\leq n+m-1}} x_ix_j
\label{clc} \enspace,\eeg
where ${\bf{x}}=(x_0,\ldots,x_{n+m-1})\in \mbox{GF}(2)^{n+m}$,
the number of flat spectra w.r.t. $\{H,N\}^{n+m}$ is as given as follows:
\begin{lem}\label{clchn}
For $n,\ m \geq1$, we have $$\begin{array}{lcl}K_{n,m}^{HN}&=&\#\mbox{ flat spectra}(p_{n,m}({\bf{x}})) \mbox{ w.r.t. } \{H,N\}^{n+m}\\
& =& 3nm-n(\frac{1+(-1)^m}{2})-m(\frac{1+(-1)^n}{2})+3(\frac{1+(-1)^n}{2})(\frac{1+(-1)^m}{2})\enspace.\end{array}$$
\end{lem}
\vspace{3mm}
\begin{proof} The generic modified adjacency matrix of the graph is as follows:

\begin{small}
$$\Gamma_{\bf{v}}=\left(\begin{array}{ccccccccccc}
v_0 & 1 & 1 & \ldots & 1 & 0 & 0 & 0 & \ldots & 0\\
1 & v_1 & 1  & \ldots & 1& 0 & 0 & 0 & \ldots & 0\\
\vdots & \vdots &  \vdots & \ddots & \vdots& \vdots & \vdots &  \vdots & \ddots & \vdots\\
1 & 1 & 1  & \ldots & v_{n-1}& 1 & 0 & 0 & \ldots & 0\\
 0 & 0 & 0 & \ldots & 1 & v_{n} & 1 & 1 & \ldots & 1\\
 0 & 0 & 0 &\ddots  & 0 & 1 & v_{n+1} & 1  & \ldots & 1\\
 \vdots & \vdots &  \vdots & \vdots & \vdots & \vdots & \vdots &  \vdots & \ddots & \vdots\\
 0 & 0 & 0 & 0 & 0 &1 & 1 & 1  & \ldots & v_{n+m-1}\end{array}\right)\enspace.$$
\end{small}

Calculating the determinant, we see that $|\Gamma_{\bf{v}}|=|G_c|+C$, where $G_c$ is the generic modified adjacency matrix of the two
independent cliques:

\begin{small}
$$G_c=\left(\begin{array}{ccccccccccc}
v_0 & 1 & 1 & \ldots & 1 & 0 & 0 & 0 & \ldots & 0\\
1 & v_1 & 1  & \ldots & 1& 0 & 0 & 0 & \ldots & 0\\
\vdots & \vdots &  \vdots & \ddots & \vdots& \vdots & \vdots &  \vdots & \ddots & \vdots\\
1 & 1 & 1  & \ldots & v_{n-1}& 0 & 0 & 0 & \ldots & 0\\
 0 & 0 & 0 & \ldots & 0 & v_{n} & 1 & 1 & \ldots & 1\\
 0 & 0 & 0 &\ddots  & 0 & 1 & v_{n+1} & 1  & \ldots & 1\\
 \vdots & \vdots &  \vdots & \vdots & \vdots & \vdots & \vdots &  \vdots & \ddots & \vdots\\
 0 & 0 & 0 & 0 & 0 &1 & 1 & 1  & \ldots & v_{n+m-1}\end{array}\right)$$
\end{small} 

and  $C$ is the product of the first $(n-1)\times(n-1)$ minor and the last $(m-1)\times(m-1)$ minor:

\begin{small}
$$C=\left|\begin{array}{cccccc}
v_0 & 1 & 1 & \ldots & 1 \\
1 & v_1 & 1  & \ldots & 1\\
\vdots & \vdots &  \vdots & \ddots & \vdots\\
1 & 1 & 1  & \ldots & v_{n-2}\end{array}\right| \cdot \left|\begin{array}{cccccc}
 v_{n+1} & 1 & 1 & \ldots & 1\\
 1 & v_{n+2} & 1  & \ldots & 1\\
 \vdots & \vdots &  \vdots & \ddots & \vdots\\
 1 & 1 & 1  & \ldots & v_{n+m-1}\end{array}\right|\enspace.$$
\end{small}

The first minor corresponds to the determinant of a clique in $n-1$ variables, say $C_1$,
and the second to that of a clique in $m-1$ variables, say $C_2$. 

 As seen in the proof of lemma \ref{lem:cliqueHN}, we have to look separately at the different cases that arise from the parities of $n$ and $m$. We will denote by $K_n^c$ the number of flat spectra of the clique in $n$ variables w.r.t. $\{H,N\}^n$.

\begin{itemize}
\item Case $n, m$ odd: Here, $C=0$ iff two or more of the $v_0,v_1,\ldots,v_{n-2}$ and/or two or more of \linebreak 
$v_{n+1},v_{n+2},\ldots,v_{n+m-1}$ are equal to 1. In that case $|G_c|=0$ as well, since there will be linear dependence in the rows of $G_c$.
Therefore the only case in which we obtain $|\Gamma_{\bf{v}}|=1$ is when $C=1$ and $|G_c|=0$.\\ 
The number of times $|C_1|=1$ is $K_{n-1}^c$, and the number of times $|C_2|=1$ is $K_{m-1}^c$.
Hence, we can have $C=1$ in $K_{n-1}^cK_{m-1}^c$ ways, and the rank of $\Gamma_{\bf{v}}$ will depend on its rows containing the variables
$v_{n-1}$ and $v_n$. The way to get $|G_c|=0$ is to make the choice of $v_{n-1}$ and $v_n$ such that it makes the first and/or second
cliques within $G_c$ not flat. Therefore,
\begin{center}$K_{n,m}^{HN}=K_{n-1}^c(2K_{m-1}^c)+K_{m-1}^c(2K_{n-1}^c-K_{n-1}^c)=3(n-1+\frac{1+(-1)^{n-1}}{2})(m-1+\frac{1+(-1)^{m-1}}{2})\enspace.$\end{center}

\item Case $n$ even, $m$ odd: Here, $C=0$ as above and also iff $v_0=v_1=\ldots=v_{n-2}=0$. In the last case it is possible to
have $|G_c|=1$ iff both cliques within $G_c$ are flat. This happens $2K_{m}^c$ times: for the first clique we have
$v_0=v_1=\ldots=v_{n-2}=0$ and so $v_{n-1}$ can be 0 or 1. Adding this to the number we found above, we get
$3(n-1+\frac{1+(-1)^{n-1}}{2})(m-1+\frac{1+(-1)^{m-1}}{2})+2m+1+(-1)^{m}\enspace.$

 \item Case $n$ odd, $m$ even: As in the previous case, we get
\begin{center}$3(n-1+\frac{1+(-1)^{n-1}}{2})(m-1+\frac{1+(-1)^{m-1}}{2})+2n+1+(-1)^{n}\enspace.$\end{center}

 \item Case $n, m$ even: In this case we have all the flat spectra of the second case, plus the number of flat spectra coming from
 $v_{n+1}=v_{n+2}=\ldots=v_{n+m-1}=0$ which are not already counted. This number is $2(K_{n-1}^c-2)$.
 Adding it to the rest we get \begin{center}$3(n-1+\frac{1+(-1)^{n-1}}{2})(m-1+\frac{1+(-1)^{m-1}}{2})+2(m+n-1)+(-1)^{m}+(-1)^{n}\enspace.$\end{center}
\end{itemize}
Summing up and simplifying, we get the desired formula.\end{proof}
Note: The formula is still valid for $n$ or $m$ equal to 1, if we consider $K_{0}^c=1$.

\subsection{Comparison}
Table I summarises our results for the $\{H,N\}^n$ transform
set. Further computational results show that, for $n \le 8$ and $n \le 5$,
the line has the maximum
number of flat spectra w.r.t. $\{H,N\}^n$ over the set of quadratics and over the
set of all
boolean functions, respectively. We can therefore conjecture the following:
\begin{cj}
Over the set of all boolean functions, the line function, as defined in (\ref{RS}), maximizes the number of flat spectra
w.r.t. $\{H,N\}^n$.
\label{cjline}
\end{cj}

\section{On the Number of Flat Spectra of Quadratic Boolean Functions with respect to $\{I,H\}^n$} \label{IH}
As in the previous section, it would also be interesting to construct boolean functions with the largest possible number of flat spectra w.r.t. $\{I,H\}^n$.
Note that for the {\em{interlace polynomial}}, $q(x)$, of a graph, as defined in \cite{Arr:Int1}, one can show that $q(1)$ is
the number of flat spectra w.r.t. $\{I,H\}^n$.

\subsection{Line}
The number of flat spectra of the line function, as defined by
(\ref{RS}), with respect to $\{I,H\}^n$, is the Fibonacci recurrence:

\begin{lem}\label{lineIH} $ K_n^{IH}=\#\mbox{ flat spectra}(p_l({\bf{x}}))
\mbox{ w.r.t. } \{I,H\}^n = K_{n-1}^{IH}+K_{n-2}^{IH} $, with $K_0^{IH} = K_1^{IH} = 1$; in closed form,
$$\frac{(1+\sqrt{5})^{n+1}+(1-\sqrt{5})^{n+1}}{2^{n+1}\sqrt{5}}\enspace.$$
\end{lem}
\vspace{3mm}
\begin{proof} 
We are first going to see that $$K(k)=
\sum_{\sum_{\lambda=0}^k t_\lambda=n-k} \prod_{j=0}^k K_{t_j}^H\enspace,$$
where $K(k)$ is the number of flat spectra when $|R_I|=k$, and
$K_i^H$ is the number of flat spectra in $i$ variables w.r.t. $\{H\}^i$. It is easy to see that $K_i^H=\frac{1+(-1)^i}{2}$, with $K_0^H=1$. 

Let $R_I=\{i_0,\ldots,i_{k-1}\}$. Then,
\beg D(k)=\mbox{det}(\Gamma_I)=D^{0,\ldots,i_0-1}D^{i_0+1,\ldots,i_1-1}\cdots D^{i_k+1,\ldots,n-1}\enspace, \label{darthvader} \eeg
where $D^{k_0,\ldots,k_t}$ is the determinant of the generic modified matrix of the line,
$\Gamma_I$, in the variables $x_{k_0},\ldots,x_{k_t}$. With a slight abuse of notation, when $i_0=0,i_{j+1}=i_{j}+1$ or $i_{k}=n-1$, we will consider the
corresponding determinant of the empty matrix to be equal to 1.
To prove this formula we use induction on $k$:

\underline{Case $k=0$} $(R_I=\emptyset)$. Evidently, $D(0)=D^{0,\ldots,n-1}$.

\underline{Case $k=1$}. In this case $R_I=\{i_0\}$. When we 'cross out' the $i_0^{th}$ row and column from the matrix,
we get a block matrix of four blocks in which both anti-diagonal blocks are zero. $D(1)=1$ if and only if the rows of the
matrix are linearly independent.
But because of the anti-diagonal blocks being zero, that happens if and only if in each of the other two blocks the rows are
linearly independent, that is if the determinants of both blocks are equal to 1. In other words, $D(1)=D^{0,\ldots,i_0-1}D^{i_0+1,\ldots,n-1}$.

Suppose the statement holds for $|R_I|=m$: if $R_I=\{j_0,\ldots,j_{m-1}\}$, $D(m)=D^{0,\ldots,j_0-1}\cdots D^{j_{m-1}+1,\ldots,n-1}$.
We will see that it is true for $|R_I|=m+1$:

Let $R_I=\{i_0,\ldots,i_{m}\}=\{j_0,\ldots, j_{l}, \lambda, j_{l+1},\ldots,j_{m-1}\}$.
Then, by induction hypothesis \linebreak
$D(m+1)=D^{0,\ldots,j_0-1}\cdots D_{\lambda}^{j_l+1,\ldots,j_{l+1}-1}\cdots D^{j_{m-1}+1,\ldots,n-1}$, where
$D_{\lambda}^{j_l+1,\ldots,j_{l+1}-1}$ represents the determinant $D^{j_l+1,\ldots,j_{l+1}-1}$ with the
$\lambda^{th}$ row and column crossed out. From the case $k=1$, we see that
$$D_{\lambda}^{j_l+1,\ldots,j_{l+1}-1}=D^{j_l+1,\ldots,\lambda-1}D^{\lambda+1,\ldots,j_{l+1}-1}\enspace,$$
and that concludes the proof of equation (\ref{darthvader}). 

The determinant on the left hand side of equation (\ref{darthvader}) is equal to 1 iff each one of the determinants on the right hand side is equal to 1.
But each determinant $D^{k_0,\ldots,k_t}$ will be equal to 1 exactly
$K_{k_t-k_0+1}^H$ times. So for $R_I=\{i_0,\ldots,i_{k-1}\}$, the number of flat
spectra is
$K_{i_0}^HK_{i_1-i_0-1}^H\cdots K_{n-1-i_{k-1}}^H$
and so
$$K(k)=\sum_{|R_I|=k}K_{i_0}^HK_{i_1-i_0-1}^H\cdots K_{n-1-i_{k-1}}^H\enspace.$$
The summands that appear in $K(k)$ are all possible products
$\prod K_i^H$ such that the sum of the indices is $n-k$, so we have
$$K(k)=\sum_{\sum_{\lambda=0}^k t_\lambda=n-k} \prod_{j=0}^k K_{t_j}^H\enspace.$$
If we write the indices as a vector, $(t_0,\ldots,t_{n-1})$,
where $\sum_{l=0}^{n-1} t_l=n-k$, then for $(t_1,\ldots,t_{n-1})$ we
have that $\sum_{l=1}^{n-1} t_l=n-k-t_0$.
Hence, for all possible vectors in
$K_n^{IH}=\sum_{k=0}^{n-1}K(k)$, we have all possible vectors in the
lesser indices, as follows:
\beg K_n^{IH}=K_n^H+K_{n-1}^HK_{0}^{IH}+K_{n-2}^HK_{1}^{IH}+\ldots+
K_{0}^HK_{n-1}^{IH}=K_n^H+\sum_{i=0}^{n-1} K_{n-1-i}^HK_{i}^{IH}\enspace.\label{rec} \eeg
For the rest of the proof,
we are going to omit the superscript $H$ and use that $K_{n}+K_{n+1}=1$.

Using (\ref{rec}), we get 
$$\begin{array}{rl}
K_{n+2}^{IH}
= &  K_{n+2} +\displaystyle \sum_{i=0}^{n+1} K_{n+1-i}K_{i}^{IH} \\[6pt]
= & K_{n+2} + \displaystyle \sum_{i=0}^n K_{n+1-i}K_{i}^{IH}+K_0K_{n+1}^{IH} \\[6pt]
= &  K_{n+2} + \displaystyle \sum_{i=0}^n K_{n+1-i}K_{i}^{IH}+K_{n+1}+\displaystyle \sum_{i=0}^{n} K_{n-i}K_{i}^{IH} \\[6pt]
= &  K_{n+1} + K_{n+2} + \displaystyle \sum_{i=0}^{n} K_{i}^{IH} (K_{n-i}+K_{n+1-i}) \\[6pt]
= &  1+\displaystyle \sum_{i=0}^{n} K_{i}^{IH}= 1 + \displaystyle \sum_{i=0}^{n-1} K_{i}^{IH} + K_{n}^{IH}\\[6pt]
= & 1 + \displaystyle \sum_{i=0}^{n-1} K_{i}^{IH}(K_{n-1-i}+K_{n-i}) +K_{n}^{IH}K_0\\[6pt]
= & K_n+K_{n+1}+\displaystyle \sum_{i=0}^{n-1} K_{n-1-i}K_{i}^{IH} + \displaystyle \sum_{i=0}^{n} K_{n-i}K_{i}^{IH}\\[6pt]
= & K_{n}^{IH}+K_{n+1}^{IH}
\end{array}$$

This gives us the recurrence relation, and from there we get
the closed formula.\end{proof}

{\bf{Remark:}} This result appears in \cite{Arr:Int1} as the evaluation
of the interlace polynomial $q(x)$ for the path graph at $x = 1$.

\subsection{Clique}
The clique function, as defined in (\ref{clique})
satisfies the following lemma:
\begin{lem}
$ K_{n}^{IH}=\#\m{ flat spectra}(p_c({\bf{x}})) \m{ w.r.t. } \{I,H\}^n = 2^{n-1}\enspace.$
\label{cliqueIH}
\end{lem}
\begin{proof}
It is easy to show from its adjacency matrix that the clique function of $n$
variables is bent for $n$ even. Consider the sub-functions of the $n$-variable
clique function, obtained by fixing a subset of the input variables, ${\bf{R_I}}$.
These sub-functions will also be cliques and will be Bent iff $n - |{\bf{R_I}}|$
is even.
The lemma follows by straightforward counting arguments.
\end{proof}

{\bf{Remark:}} This result appears in \cite{Arr:Int1} as the evaluation
of the interlace polynomial $q(x)$ for the complete graph at $x = 1$.

\subsection{Clique-Line-Clique}
For the $n$ clique-line-$m$ clique, as defined in (\ref{clc}), we  get:
\begin{lem} For $n, \ m \geq 1$ such that $n+m\geq4$,\\
$ K_{n,m}^{IH}=\#\mbox{ flat spectra}(p_{n,m}({\bf{x}})) \mbox{ w.r.t. } \{I,H\}^{n+m} =
2K_{n-1,m}^{IH}=2K_{n,m-1}^{IH}$; in closed form,
$$K_{n,m}^{IH}=5\cdot2^{n+m-4}\enspace.$$ \end{lem}
\vspace{3mm}
\begin{proof}We begin the proof with some observations. Firstly, note that by fixing one of the connecting variables, $x_{n-1}$ or $x_n$, we get two independent cliques, either in $n-1$ and $m$ variables respectively or in $n$ and $m-1$ variables respectively. Secondly, if we fix any of the other variables instead,
we get the same kind of clique-line-clique graph. Thirdly, from the proof of lemma \ref{clchn}, we can deduce that $p_{n,m}$ is bent iff $n+m$ is even.

By the first and second observations, and considering that the order in which we fix doesn't matter, we get three separate cases:
\begin{itemize}
\item Case 1: We fix any variables but the connecting ones. Then, by the second and third observations, we have flat
spectra by fixing $t$ variables iff $n+m-2-t$ is even; that is, if $n+m-t$ is even.
Therefore the number of flat spectra for this case is:
$$ \begin{small}
N1= \left\{\begin{array}{l l}
\displaystyle\sum_{k=0}^{(n+m)/2}\left(\begin{array}{c}
n+m-2\\
2k\end{array}\right)&\mbox{if} \ n+m\mbox{ even}\\[6pt]
\displaystyle\sum_{k=0}^{(n+m-1)/2}\left(\begin{array}{c}
n+m-2\\
2k+1\end{array}\right)&\mbox{if} \ n+m\mbox{ odd}
\end{array}\right. \end{small} $$
\item Case 2: We fix $x_{n-1}$. We thus have two independent cliques, one of $n-1$  and the other of $m$ variables.
We can then fix any of the remaining variables; when we fix $t_1$ variables in the first clique and $t_2$ in the second, we obtain a flat spectrum iff $n-1-t_1$ and $m-t_2$ are both even.
Thus,
$$N2=2^{n-2}2^{m-1}\enspace.$$
\item Case 3: We fix $x_n$. We then get two independent cliques, one of $n$ and the other of $m-1$ variables.
We can now fix any of the remaining variables but $x_{n-1}$; when we fix $t_1$ variables in the first clique and $t_2$ in the second, we obtain a flat spectrum iff $n-t_1$ and $m-1-t_2$ are both even. Thus,
$$ \begin{small} N3=2^{m-2}\cdot\left\{\begin{array}{l l}
\displaystyle\sum_{k=0}^{(n-1)/2}\left(\begin{array}{c}
n-1\\
2k\end{array}\right) &\mbox{if}\ n\mbox{ odd}\\[6pt]
\displaystyle\sum_{k=0}^{(n-2)/2}\left(\begin{array}{c}
n-1\\
2k+1\end{array}\right)&\mbox{if}\ n\mbox{ even}
\end{array}\right. \end{small} $$
\end{itemize}

Clearly, $K_{n,m}^{IH}=N1+N2+N3$; in principle, the result depends on the parity of $n$ and $m$. However
$$ \begin{small} \displaystyle\sum_{k=0}^{s/2}\left(\begin{array}{c}
s\\
2k\end{array}\right)=1+\displaystyle\sum_{k=1}^{s/2}\left[\left(\begin{array}{c}
s-1\\
2k\end{array}\right)+\left(\begin{array}{c}
s-1\\
2k-1\end{array}\right)\right]=1+\displaystyle\sum_{i=1}^{s-1}\left(\begin{array}{c}
s-1\\
i\end{array}\right)=2^{s-1} \end{small} \enspace,$$
and in the same way
$$ \begin{small} \displaystyle\sum_{k=0}^{(s-1)/2}\left(\begin{array}{c}
s\\
2k+1\end{array}\right)=\displaystyle\sum_{k=1}^{(s-1)/2}\left[\left(\begin{array}{c}
s-1\\
2k+1\end{array}\right)+\left(\begin{array}{c}
s-1\\
2k\end{array}\right)\right]=\displaystyle\sum_{i=0}^{s-1}\left(\begin{array}{c}
s-1\\
i\end{array}\right)=2^{s-1} \end{small} \enspace.$$

Therefore, in all cases, we get $K_{n,m}^{IH}=N1+N2+N3=5\cdot2^{n+m-4}$, and from here,
trivially, the recurrence relation.
\end{proof}

\subsection{Comparison}
Table I summarises our results for the $\{I,H\}^n$ transform
set. As seen from both our theoretical and computational results, the clique function
has the maximum number of flat spectra w.r.t. $\{I,H\}^n$ over the set of
quadratics for $n \le 8$, and over the set of all boolean functions for $n \le 5$. Hence we can conjecture the following:
\begin{cj}
Over the set of all boolean functions,
the clique function, as defined in (\ref{clique}), maximises the number of
flat spectra w.r.t. $\{I,H\}^n$.
\label{cjclique}
\end{cj}

\section{On the Number of Flat Spectra of Boolean Functions with respect to $\{I,H,N\}^n$} \label{IHN}
As deduced from computational results, high-distance stabilizer
quantum codes (optimal additive codes over $\mbox{GF}(4)$) are
associated to quadratic boolean functions with large numbers of
flat spectra w.r.t. $\{I,H,N\}^n$. In fact Hein et al \cite{Hein:GrEnt} have already argued that
high-distance QECCs will represent highly-entangled pure multipartite quantum states, and one indication of this entanglement
strength will be an 'evenly-spread' power spectrum  w.r.t. all {\em{Local Unitary Transforms}} \cite{Par:QE}, of which
$\{I,H,N\}^n$ is a strategic subset.
Therefore, the problem of maximising the number of flat spectra
w.r.t. $\{I,H,N\}^n$ is of
significant importance. As a means of comparison, we first consider
the number of flat spectra for the near-worst and worst-case functions,
namely the constant function and the monomial function of degree $n$,
respectively.

\subsection{Constant function}
\begin{lem}
The {\em constant function} in $n$ variables, $p({\bf{x}}) = 0$ or $1$, where ${\bf{x}}=(x_0,\ldots,x_{n-1})\in \mbox{GF}(2)^n$, has $2^n$ flat spectra with respect to $\{I,H,N\}^n$.
\end{lem}
\begin{proof}
Any $\{I,N\}^n$
transform of the constant function is flat, and none of the others: as seen in \cite{RP:BC}, we get flat spectra iff  $p_I({\bf{x}}) + p_I({\bf{x}}+ {\bf{k}})+ \sum_{i=1}^{n-1} \chi_{_{\bf{R_N}}}(i)k_i x_i$  is balanced for all ${\bf{k}}\neq{\bf{0}}$, where ${\bf{k}}=(k_0,\ldots,k_{n-1})\in \mbox{GF}(2)^n$, $\chi_{_{\bf{R_N}}}$ is the characteristic function of the set ${\bf{R_N}}$ and $p_I$ is the restriction of the function when fixing the variables whose indices are in ${\bf{R_I}}$. In our case, for any choice of ${\bf{R_I}}$, we get $p_I({\bf{x}})+p_I({\bf{x}}+{\bf{k}})=0$.
Thus, we get flat spectra iff $\sum_{i=0}^{n-1} \chi_{_{\bf{R_N}}}(i)k_i x_i$ is balanced for all ${\bf{k}}\neq{\bf{0}}$. Clearly, if $\chi_{_{\bf{R_N}}}(i)=1$ for all $i\in \{0,\ldots,n-1\}\setminus {\bf{R_I}}$, we get a balanced function for all ${\bf{k}}\neq{\bf{0}}$. But if $i\in {\bf{R_H}}$ for some $i$, $\chi_{_{\bf{R_N}}}(i)=0$, and by taking ${\bf{k}}=(0,\ldots,1,\ldots,0)$, where the 1 is in the $i^{th}$ position, we get an unbalanced function.
\end{proof}

\subsection{Monomial function}
\begin{lem}
The {\em monomial function} of degree $n$ in $n$ variables,
$p({\bf{x}}) = x_0x_1x_2\ldots x_{n-1}$, where \linebreak ${\bf{x}}=(x_0,\ldots,x_{n-1})\in \mbox{GF}(2)^n$, has
$n + 1$ flat spectra w.r.t. $\{I,H,N\}^n$, except for the case $n=2$.\end{lem}

\begin{proof}
Throughout this proof, we will use the same notation as in the previous one.

We first let $n=1$. Then, the monomial function becomes the linear function $x_0$
in one variable. This will have the same flat spectra as the constant
function in one variable, that is $2^1=n+1$.

Next, we let $n=2$. Then the monomial is the same as the line in two variables, and will be considered in lemma \ref{lineIHN}.

Now, we let $n>2$ and ${\bf{R}}=\{i_0,\ldots,i_l\}=\{0,\ldots,n-1\}\setminus {\bf{R_I}}$. Suppose that we fix $x_i=1$ for all $i\ \in\ {\bf{R_I}}$, and that $|{\bf{R}}|>2$. If we take ${\bf{k}}=(1,0,\ldots,0)$, the function $p_I({\bf{x}}) + p_I({\bf{x}}+ {\bf{k}})
+ \sum_{i=0}^{n-1} \chi_{_{\bf{R_N}}}(i)k_i x_i$ becomes $x_{i_1}\cdots x_{i_l}+\chi_{_{\bf{R_N}}}(i_0)x_{i_0}$,
which is balanced iff $\chi_{_{\bf{R_N}}}(i_0)=1$.
Similarly, we see that we must have $\chi_{_{\bf{R_N}}}(i)=1$ for all $i\ \in\ {\bf{R}}$ (that is, ${\bf{R}}={\bf{R_N}}$).
Consider now ${\bf{k}}=(1,1,0\ldots,0)$. The function we will get is
$x_{i_1}\cdots x_{i_l}+x_{i_0}x_{i_2}\cdots x_{i_l}+x_{i_2}\cdots x_{l}+x_{i_0}+x_{i_1}$,
which is not balanced. 

Therefore, for $n>2$, we need to fix at least $n-2$ variables in order
to obtain flat spectra; that is, we need $|{\bf{R_I}}|\geq n-2$.
Suppose now $|{\bf{R_I}}|= n-2$: By symmetry, we can suppose, w.l.o.g.,
that we fix $x_2,\ldots,x_{n-1}$. If any of the $x_i=0$, then
our new function is a constant, $p_I=0$. As we have just seen, the only
possibility for
$p_I({\bf{x}}) + p_I({\bf{x}}+ {\bf{k}})+ \chi_{_{\bf{R_N}}}(0)k_0 x_0+\chi_{_{\bf{R_N}}}(1)k_1 x_1$
to be balanced for all ${\bf{k}}\neq(0,0)$ is that
$\chi_{_{\bf{R_N}}}(0)=\chi_{_{\bf{R_N}}}(1)=1$.
On the other hand, if $x_i=1$ for all $i\geq 2$, $p_I=x_0x_1$,
the line in two variables; as we can easily deduce from the generic modified
adjacency matrix, it has a flat spectrum iff  $\chi_{_{\bf{R_N}}}(i)=0$ for at least one of the $i$'s.
Thus we get a contradiction, and so in fact $|{\bf{R_I}}|\geq n-1$.
When $|{\bf{R_I}}|=n-1$, by fixing we now get either $p_I=0$ or $p_I=x_i$.
Both have a flat spectrum iff $\chi_{_{\bf{R_N}}}(i)=1$, and from here we
get $n$ flat spectra. Finally, for $|{\bf{R_I}}|=n$, we get another flat spectrum.\end{proof}

{\bf{Remark:}} It can be shown that $n+1$ is the minimal number of flat spectra possible for a boolean function
w.r.t. $\{I,H,N\}^n$.

\subsection{Line}
As opposed to the case of $\{H,N\}^n$, the number of flat spectra of the line w.r.t. $\{I,H,N\}^n$ does not seem to be maximal:

\begin{lem}\label{lineIHN} $ K_n^{IHN}=\#\mbox{ flat spectra}(p_l({\bf{x}})) \mbox{ w.r.t. }
\{I,H,N\}^n = 2(K_{n-1}^{IHN}+K_{n-2}^{IHN}) $, with $K_0^{IHN}= 1$ and $K_1^{IHN}= 2$; in closed form,
$$K_n^{IHN}=
\frac{(1+\sqrt{3})^{n+1}-(1-\sqrt{3})^{n+1}}{2\sqrt{3}}\enspace.$$
 \end{lem}
\vspace{3mm}
\begin{proof}
Following the same arguments as in the proof of
Lemma \ref{lineIH}, we arrive at the formula:
\beg K_n^{IHN}=K_n+\sum_{i=0}^{n-1} K_{n-1-i}K_{i}^{IHN}\enspace, \label{Jedi} \eeg
where here, $K_i$ will represent the number of flat spectra in $i$ variables w.r.t. $\{H, N\}^n$.

In the sequel we are going to use that
$K_n=2^n-K_{n-1}$ (see Lemma \ref{lem:lineHN}),
or more accurately its consequence
$K_{n+1}+K_{n+2}=2^{n+2}=2(K_{n}+K_{n+1})$. We will also use that $K_0=K_1=1$.

Using (\ref{Jedi}) we get 
$$\begin{array}{l c l} 
2K_{n}^{IHN}+2K_{n+1}^{IHN} & = & 2K_n+2K_{n+1}+2\displaystyle \sum_{i=0}^{n-1} K_{n-1-i}K_{i}^{IHN}+2 \displaystyle \sum_{i=0}^{n} K_{n-i}K_{i}^{IHN}\\
& = &K_{n+2}+K_{n+1}+ \displaystyle \sum_{i=0}^{n-1} K_{i}^{IHN}2(K_{n-1-i}+K_{n-i}) +2K_{n}^{IHN}K_0\\
& = & K_{n+2}+\displaystyle \sum_{i=0}^{n-1} K_{i}^{IHN}(K_{n-i}+K_{n-i+1}) +K_{n}^{IHN}(K_0+K_1)+K_{n+1}\\
& = &K_{n+2}+\displaystyle \sum_{i=0}^{n} K_{i}^{IHN}(K_{n-i}+K_{n-i+1})+K_{n+1}\\
& = &K_{n+2}+\displaystyle \sum_{i=0}^{n} K_{i}^{IHN}K_{n-i+1} + K_{n+1}+ \displaystyle \sum_{i=0}^{n} K_{i}^{IHN}K_{n-i}\\
& = &K_{n+2}+\displaystyle \sum_{i=0}^{n} K_{i}^{IHN}K_{n-i+1}+K_{n+1}^{IHN}\\
& = &K_{n+2}+\displaystyle \sum_{i=0}^{n+1} K_{i}^{IHN}K_{n-i+1}=K_{n+2}^{IHN}
\end{array}$$
From here, we arrive to the closed formula. \end{proof}

{\bf{Remark:}} This result can be gleaned, indirectly, from page 23 of
\cite{Aig:Int} as the evaluation
of the interlace polynomial $Q(x)$ for the path graph at $x = 2$.

\subsection{Clique}
Although the clique function as defined in (\ref{clique}) appears to be maximal
w.r.t. $\{I,H\}^n$, it does not do so well w.r.t.
$\{I,H,N\}^n$: 

\begin{lem}
$ K_n^{IHN}=\#\mbox{ flat spectra}(p_c({\bf{x}})) \mbox{ w.r.t. } \{I,H,N\}^n = 2K_{n-1}^{IHN}+2^n$;
in closed form, $$K_n^{IHN}=(n+1)2^{n-1}\enspace.$$ \end{lem}
\vspace{3mm}
\begin{proof} As stated before, if we have a clique in $n$ variables and we fix a subset in the set of variables (that is, we choose ${\bf R_I}$), we get a clique in $n-|{\bf R_I}|$ variables.
Thereby, for each selection of ${\bf R_I}$ we have as many flat spectra as the number of flat spectra w.r.t. $\{H,N\}^{n-|{\bf R_I}|}$,
in $n-|{\bf R_I}|$ variables. Therefore,
$$ \#\mbox{ flat spectra}(p_c({\bf{x}})) \mbox{ w.r.t. }
\{I,H,N\}^n = \sum_{i=0}^n \left(\begin{array}{c}
n\\i\end{array}\right)K_{n-i}\enspace,$$ where $K_{n-i}$ is the number of flat spectra of the clique in $n-i$ variables w.r.t. $\{H,N\}^{n-i}$. Now, 
\begin{small}
$$\begin{array}{lcl}
\displaystyle \sum_{i=0}^n \left(\begin{array}{c}n\\i\end{array}\right)K_{n-i} 
& = &\displaystyle \sum_{i=0}^n \left(\begin{array}{c}n\\n-i\end{array}\right)K_{i}= \displaystyle \sum_{i=0}^n 
\left(\begin{array}{c}n\\i\end{array}\right)K_{i}\\ [12pt]
& = &\displaystyle \sum_{i=0}^n \left(\begin{array}{c}n\\i\end{array}\right)\left(i+\frac{1+(-1)^i}{2}\right)\\ [12pt]
& = &\displaystyle \sum_{i=0}^n \left(\begin{array}{c}n\\i\end{array}\right)i+\displaystyle \sum_{i=0}^n 
\left(\begin{array}{c}n\\i\end{array}\right)\frac{1}{2}+\displaystyle \sum_{i=0}^n 
\left(\begin{array}{c}n\\i\end{array}\right)\frac{(-1)^i}{2}\\ [12pt]
& = &\displaystyle \sum_{i=0}^n \left(\begin{array}{c}n\\i\end{array}\right)i+2^{n-1}+0\end{array}$$
\end{small}
Expanding the first term,
$$ \begin{small} 
\begin{array}{lcl}
\displaystyle \sum_{i=0}^n \left(\begin{array}{c}n\\i\end{array}\right)i
& = &\left(\begin{array}{c}n\\0\end{array}\right)0+\left(\begin{array}{c}
n\\n\end{array}\right)n+\displaystyle \sum_{i=1}^{n-1} \left(\begin{array}{c}
n\\i\end{array}\right)i\\ [12pt]
& = & n+ \ \displaystyle \sum_{i=1}^{n-1} \left[\left(\begin{array}{c}
n-1\\i\end{array}\right)+\left(\begin{array}{c}
n-1\\i-1\end{array}\right)\right]i\\ [12pt]
& = & n+ \ \displaystyle \sum_{i=0}^{n-1} \left(\begin{array}{c}
n-1\\i\end{array}\right)i+\displaystyle \sum_{i=0}^{n-1} \left(\begin{array}{c}
n-1\\i-1\end{array}\right)i\\ [12pt]
& = &n+2\displaystyle \sum_{i=0}^{n-1} \left(\begin{array}{c}
n-1\\i\end{array}\right)i-\left(\begin{array}{c}
n-1\\n-1\end{array}\right)(n-1)+\displaystyle \sum_{i=0}^{n-2} \left(\begin{array}{c}
n-1\\i\end{array}\right)\\ [12pt]
& = & \ \ 2 \displaystyle \sum_{i=0}^{n-1} \left(\begin{array}{c}
n-1\\i\end{array}\right)+1+2^{n-1}- \left(\begin{array}{c}
n-1\\n-1\end{array}\right)\end{array}\end{small} $$ 
\vspace{2mm}
Hence, we get that
$K_n^{IHN}=2K_{n-1}^{IHN}+2^n$. From the recurrence relation we get the
desired formula.\end{proof}

{\bf{Remark:}} For the cases $n = 2, \ 3,$ and $4, \ K_n^{IHN}$ of the clique function can be found
by evaluating the interlace polynomial $Q(x)$ for the complete graph at $x = 2$ (\cite{Aig:Int}, p.21).

\subsection{Clique-Line-Clique}
For the $n$ clique-line-$m$ clique structure, as defined in (\ref{clc}), the number of flat spectra is as follows:

\begin{lem} $ K_{n,m}^{IHN}=\#\mbox{ flat spectra}(p_{n,m}({\bf{x}})) \mbox{ w.r.t. } \{I,H,N\}^{n+m} = 2^{n+m-3}(3nm+2n+2m+2)\enspace.$ \end{lem}

\begin{proof} Suppose that one or both of the connecting variables are in ${\bf{R_I}}$: when we fix one of the connecting variables,
we get two independent cliques, so from this case we get
$$K_{n-1,C}^{IHN}K_{m,C}^{IHN}+K_{n,C}^{IHN}K_{m-1,C}^{IHN}-K_{n-1,C}^{IHN}K_{m-1,C}^{IHN}=2^{m+n-4}(3nm+2n+2m)\enspace,$$
where $K_{k,C}^{IHN}$ is the number of flat spectra of the clique in $k$ variables w.r.t. $\{I,H,N\}^k$.

On the other hand, when none of the connecting variables are in ${\bf{R_I}}$, we get another clique-line-clique: suppose that we
fix $i$ variables in the first clique and $j$ in the second one. In that case, we will have as many flat spectra as the number of flat spectra w.r.t. $\{H,N\}^{n+m-i-j}$ of an $(n-i)$ clique-line-$(m-j)$ clique. Considering all possible fixings in this case, we get:

$$\displaystyle \sum_{i=0}^{n-1}\displaystyle \sum_{j=0}^{m-1}\begin{small}\left(\begin{array}{c}
n-1\\i\end{array}\right)\left(\begin{array}{c}m-1\\j\end{array}\right)\end{small}K_{n-i,m-j}^{HN}
=2^{m+n-4}(3nm+2n+2m+4)\enspace.$$

\end{proof}

\subsection{Comparison}
It is well-known that
optimal $\GF(4)$-additive codes make optimal QECCs \cite{Cald:Qua}.
The mapping from a quadratic boolean function to a $\GF(4)$-additive code
is as follows. Let $p({\bf{x}})$ be a quadratic function over $n$ variables
with associated adjacency matrix, $\Gamma$. Then the generator matrix for
a $[n,2^n,d]$ $\GF(4)$-additive code is given by $\Gamma + \omega I_n$,
where $\omega^2 + \omega + 1 = 0$ over $\GF(4)$ and $I_n$ is the $n \times n$
identity matrix. This $\GF(4)$-additive code can be interpreted as a $[[n,0,d]]$
QECC of the stabilizer type.
Using the database at \cite{Dan:Dat}, an
exhaustive computer search for $n$ variable quadratic boolean functions, $4\leq n\leq 9$, finds one unique
Local complementation (LC) orbit of functions for each $n$, whose number of flat
spectra with respect to $\{I,H,N\}^n$ is optimal. A representative
for each of these orbits is listed in Table II.
All of these functions map to additive zero-dimension QECCs with optimal distances (see \cite{Gras:QECCs} and \cite{Dan:Dat}). 
 
It remains open as to whether
the quadratic function with the optimal
number of flat spectra w.r.t. $\{I,H,N\}^n$ will always have optimal distance
when viewed as a QECC, and vice versa.
In any case, the approximate correspondence is to be expected
as the QECC distance is equal to the
{\em{aperiodic propagation criteria (APC) distance}} of the
quadratic boolean functions, as presented in \cite{DanAPC}. Furthermore, optimal
propagation (aperiodic autocorrelation) criteria will relate to very
good spectral properties via a generalised form of Fourier duality.

Tables III to V show an exhaustive computer search for
boolean functions that achieve the optimal number of flat spectra
w.r.t. $\{I,H,N\}^n$ for cubics, quartics, and quintics respectively, where 
one representative function is given per LC orbit. As
expected, the maximum number of flat spectra decreases as the algebraic degree
of the boolean function rises. Also shown is the distance of the boolean
function when viewed as a zero-dimensional (non-stabilizer) QECC. As with the quadratics, this
distance parameter can be interpreted as the APC distance
of a boolean function (see \cite{DanAPC} for more details). In all cases, the
boolean functions shown in the tables achieve the maximum possible distance for
their given algebraic degree.

\section{Conclusion} \label{conc}
We derived simple recursions for the number of flat spectra with respect to
$\{I,H,N\}^n$ for certain recursive
quadratic boolean constructions, and we demonstrated that Quantum Error
Correcting Codes with optimal distance appear to have the most flat spectra
with respect to $\{I,H,N\}^n$, at least for small $n$. In subsequent work we hope
to develop recursive formulae for nested-clique structures of the type highlighted
in \cite{DanPAR}, as we expect that these will have many flat spectra w.r.t. $\{I,H,N\}^n$.

We also showed computationally that,
for small $n$, the number of flat spectra decreases when the algebraic
degree of the boolean function increases. Future work should
seek to establish constructions for boolean functions of degree greater than
two that have as large a number of flat spectra as possible
w.r.t. $\{I,H,N\}^n$.
More generally, it would
be of interest to relax the criteria somewhat, and look for those functions
which have many spectra with respect to $\{I,H,N\}^n$ with a worst-case
spectral power peak less than some low upper bound (see \cite{DanPAR}). One would expect,
in this case, that many more boolean functions of degree $ > 2$ would be found
that do well for this relaxed criteria. One promising line of inquiry in this context would
be to apply and specialise the construction proposed at the end of \cite{DanPAR}, which takes
a global graph structure, where the graph 'nodes' partition the set of boolean variables, and
where the nodes are 'linked' by permutations over these variable subsets, thereby obtaining
higher-degree boolean functions with potentially favourable $\{I,H,N\}^n$ spectra.

Finally we have answered, indirectly, a question posed at the
end of \cite{Arr:Int1} as to a simple combinatorial explanation of the
interlace polynomial $q$. It is evident that $q$ summarises some of the
spectral properties of the graph w.r.t. $\{I,H\}^n$.
Similarly the interlace polynomial $Q$, as defined in \cite{Aig:Int},
summarises some of
the spectral properties of the graph w.r.t. $\{I,H,N\}^n$.
Furthermore our work provides a natural setting for future investigations
into the generalisation of the interlace polynomial to hypergraphs.


\newpage

\section{Appendix} \label{tables}
\begin{table}[htb]
\cb
\begin{small}
\begin{tabular}{|c|c|c|c|c|c|} \hline
$\!\!\!\!\!\!\!\!\!\!\!$Function$\!\!\!\!\!\!\!\!\!$ & $\!\!\!\!\!\!\!\!\!$\begin{tabular}{c}Monomial\\ ($n>2$)\end{tabular}$\!\!\!\!\!\!\!\!\!$ & $\!\!\!\!\!$ Constant $\!\!\!\!\!\!$ &
$\!\!$ Line $\!\!$ & $\!\!$ Clique $\!\!$ & $\!\!$ $n$ clique-Line-$m$ clique $\!\!$ \\  \hline \hline
$\!\!$ ANF   & $\!\!\! x_0\ldots x_{n-1}\!\!\!$        &  0   & $\!\!\displaystyle \sum_{j=0}^{n-2} x_jx_{j+1}\!\!$   &  $\!\!\displaystyle \sum_{_{0\leq i < j\leq n-1}} x_ix_j \!\! $ &
$\displaystyle \sum_{_{0\leq i < j\leq n-1}} x_ix_j+x_{n-1}x_{n} + \sum_{_{n\leq i < j\leq n+m-1}} x_ix_j $ \\ 
 & & & & & \\
$\!\! K_n^{HN}(K_{n,m}^{HN})\!\!$   & 0  &  $2^n$  & $\frac{1}{3}\left(2^{n+1}+(-1)^n \right) $ &  $ n+\frac{1+(-1)^n}{2}$ & $\begin{array}{c}3nm-n(\frac{1+(-1)^m}{2})-m(\frac{1+(-1)^n}{2})\\ +3(\frac{1+(-1)^n}{2})(\frac{1+(-1)^m}{2})\end{array}$\\
& & & & & \\
$\!\!K_n^{IH}(K_{n,m}^{IH})\!\!$  & 1 & 1 & $\!\!\frac{(1+\sqrt{5})^{n+1}+(1-\sqrt{5})^{n+1}}{2^{n+1}\sqrt{5}}\!\!$  &  $ 2^{n-1}$ & $5\cdot2^{n+m-4}$ \\
& & & & & \\
$\!\!\!K_n^{IHN}(K_{n,m}^{IHN})\!\!\!$  &  $n + 1$ & $2^n$ & $\!\!\frac{(1+\sqrt{3})^{n+1}-(1-\sqrt{3})^{n+1}}{2\sqrt{3}}\!\!$ &  $(n+1)2^{n-1}$ & $2^{n+m-3}(3nm+2n+2m+2)$\\
& & & & & \\\hline
\end{tabular}
\caption{The Number of Flat Spectra w.r.t. $\{H,N\}^n$, $\{I,H\}^n$, and $\{I,H,N\}^n$ for some Quadratic Boolean Functions}
\end{small}
\label{flatsummary}
\ce
\end{table}

 
\begin{table}[htb]
\cb
\begin{small}
\begin{tabular}{|c|c|c|c|c|} \hline
$n$ & distance &      Quadratics Optimal for $K_n^{IHN}$ & $K_n^{IHN}$ & $K_n^{IHN}$ for the line \\ \hline \hline
4   & 2        &  02,13,23                                & 44      &  44 \\
5   & 3        &  01,02,13,24,34                          & 132     &  120 \\
6   & 4        & 01,02,05,13,15,24,25,34,35,45            & 396     &  328 \\
7   & 3        & 03,06,14,16,25,26,34,35,45               & 1096    &  896 \\
8   & 4        & 02,03,04,12,13,15,26,37,46,47,56,57,67   & 3256    &  2448 \\
9   & 4        & 04,07,08,14,16,18,25,26,28,34,35,37,57,58,67,68 & 9432    &  6688 \\ \hline
\end{tabular}
\caption{The Maximum Number of Flat Spectra w.r.t. $\{I,H,N\}^n$ for Quadratic Boolean Functions}
\end{small}
\label{flatmax}
\ce
\end{table}


\begin{table}[htb]
\cb
\begin{small}
\begin{tabular}{|c|c|c|c|} \hline
$n$ & distance &  Cubics Optimal for $K_n^{IHN}$         & $K_n^{IHN}$  \\ \hline \hline
3   &   1      &  012                                     & 4           \\
4   &   2      &  012,03,13,23                            & 20          \\
5   &   2      &  012,03,14,23,24                         & 72         \\
6   &   3      &  012,03,04,13,15,24,25                   & 248         \\ \hline
\end{tabular}
\caption{The Maximum Number of Flat Spectra w.r.t. $\{I,H,N\}^n$ for Cubic Boolean Functions}
\end{small}
\label{flatmax3}
\ce
\end{table}


\begin{table}[htb]
\cb
\begin{small}
\begin{tabular}{|c|c|c|c|} \hline
$n$ & distance &       Quartics Optimal for $K_n^{IHN}$  & $K_n^{IHN}$  \\ \hline \hline
4   &   1      &  \m{All Quartics}                        & 5          \\
5   &   2      &  0123,01,04,14,23,24,34                  & 30 \\
    &          &  0123,02,04,13,14,23,24,34               &    \\
    &          &  0123,04,14,23,24,34                     &    \\ \hline
\end{tabular}
\caption{The Maximum Number of Flat Spectra w.r.t. $\{I,H,N\}^n$ for Quartic Boolean Functions}
\end{small}
\label{flatmax4}
\ce
\end{table}


\begin{table}[htb]
\cb
\begin{small}
\begin{tabular}{|c|c|c|c|} \hline
$n$ & distance &       Quintics Optimal for $K_n^{IHN}$  & $K_n^{IHN}$  \\ \hline \hline
5   &  1       &  \m{All Quintics}   & 6  \\ \hline
\end{tabular}
\caption{The Maximum Number of Flat Spectra w.r.t. $\{I,H,N\}^n$ for Quintic Boolean Functions}
\end{small}
\label{flatmax5}
\ce
\end{table}


\begin{thebibliography}{99}

\bibitem{Aig:Int}
M. Aigner and H. van der Holst,
{"Interlace Polynomials",}
{\em Linear Algebra and its Applications},
{\bf 377}, pp. 11--30, 2004.

\bibitem{Arr:Int}
R. Arratia, B. Bollobas, and G.B. Sorkin,
{"The Interlace Polynomial: a new graph polynomial",}
{\em Proc. 11th Annual ACM-SIAM Symp. on Discrete Math.},
pp. 237--245, 2000.

\bibitem{Arr:Int1}
R. Arratia, B. Bollobas, and G.B. Sorkin,
{"The Interlace Polynomial of a Graph",}
{\em J. Combin. Theory Ser. B},
{\bf{92}}, 2, pp. 199--233, 2004.
Preprint:
\href{http://arxiv.org/abs/math/0209045}
{http://arxiv.org/abs/math/0209045},
v2, 13 Aug. 2004.

\bibitem{Arr:Int2}
R. Arratia, B. Bollobas, and G.B. Sorkin,
{"Two-Variable Interlace Polynomial",}
{\em Combinatorica},
{\bf{24}}, 4, pp. 567--584, 2004.
Preprint:
\href{http://arxiv.org/abs/math/0209054}
{http://arxiv.org/abs/math/0209054},
v3, 13 Aug. 2004.

\bibitem{Bou:Mart}
A. Bouchet,
{"Tutte-Martin Polynomials and Orienting Vectors of Isotropic Systems",}
{\em Graphs Combin.},
{\bf 7}, pp. 235--252, 1991.

\bibitem{Cald:Qua}
A.R. Calderbank,E.M. Rains,P.W. Shor and N.J.A. Sloane,
{"Quantum Error Correction Via Codes Over $\GF(4)$,"}
{\em IEEE Trans. on Inform. Theory,}
{\bf 44}, pp. 1369--1387, 1998,
(preprint:
\href{http://xxx.soton.ac.uk/abs/quant-ph/?9608006}
{http://xxx.soton.ac.uk/abs/quant-ph/?9608006}).

\bibitem{Dan:Dat}
L.E. Danielsen,
{"Database of Self-Dual Quantum Codes"},
\href{http://www.ii.uib.no/~larsed/vncorbits/}
{\it{http://www.ii.uib.no/\~{}larsed/vncorbits/}},
2004.

\bibitem{DanAPC}
L.E. Danielsen, T.A. Gulliver and M.G. Parker,
{"Aperiodic Propagation Criteria for Boolean Functions,"}
{\em ECRYPT Document Number: STVL-UiB-1-APC-1.0, submitted to Inform. Comp.},
\href{http://www.ii.uib.no/~matthew/GenDiff4.pdf}
{http://www.ii.uib.no/\~{}matthew/GenDiff4.ps},
August 2004.

\bibitem{DanPAR}
L.E. Danielsen and M.G. Parker,
{"Spectral Orbits and Peak-to-Average Power Ratio
of Boolean Functions with respect to the $\{I,H,N\}^n$ Transform",}
{\em SETA'04, Sequences and their Applications, Seoul, Accepted for
Proceedings of SETA04, Lecture Notes in Computer Science, Springer-Verlag, 2005},
\href{http://www.ii.uib.no/~matthew/seta04-parihn.pdf}
{http://www.ii.uib.no/\~{}matthew/seta04-parihn.ps},
October, 2004.

\bibitem{Dav:PF}
J.A. Davis and J. Jedwab,
{"Peak-to-mean Power Control in OFDM, Golay Complementary Sequences and Reed-Muller
Codes,"}
{\em IEEE Trans. Inform. Theory},
Vol 45, No 7, pp. 2397--2417, Nov 1999.

\bibitem{Gol:Comp}
M.J.E. Golay,
{"Complementary Series",}
{\em IRE Trans. Inform. Theory},
{\bf IT-7}, pp. 82--87, Apr. 1961.

\bibitem{Gras:QECCs}
M. Grassl,
{"Bounds on dmin for additive $[[n,k,d]]$ QECC"},
\href{http://iaks-www.ira.uka.de/home/grassl/QECC/TableIII.html}
{http://iaks-www.ira.uka.de/home/grassl/QECC/TableIII.html},
Feb. 2003.

\bibitem{Hein:GrEnt}
M. Hein, J. Eisert and H.J. Briegel,
{"Multi-Party Entanglement in Graph States",}
{\em Phys. Rev. A},
{\bf 69}, 6, 2004.
Preprint: \href{http://xxx.soton.ac.uk/abs/quant-ph/0307130}
{http://xxx.soton.ac.uk/abs/quant-ph/0307130}.

\bibitem{Par:QE}
M.G. Parker and V. Rijmen,
{"The Quantum Entanglement of Binary and Bipolar Sequences",}
short version in {\em Sequences and Their Applications},
Discrete Mathematics and
Theoretical Computer Science Series, Springer-Verlag, 2001,
long version at
\href{http://xxx.soton.ac.uk/abs/quant-ph/?0107106}
{http://xxx.soton.ac.uk/abs/quant-ph/?0107106}
or
\href{http://www.ii.uib.no/~matthew/BergDM2.ps}
{http://www.ii.uib.no/\~{}matthew/BergDM2.ps},
June 2001.

\bibitem{Par:LowPAR}
M.G. Parker and C. Tellambura,
"A Construction for Binary Sequence Sets with Low Peak-to-Average Power Ratio",
{\em Technical Report No 242, Dept. of Informatics,
University of Bergen, Norway},
\href{http://www.ii.uib.no/publikasjoner/texrap/ps/2003-242.ps}
{http://www.ii.uib.no/publikasjoner/texrap/ps/2003-242.ps},
Feb 2003.

\bibitem{RP:BC}
C. Riera and M.G. Parker,
{"Generalised Bent Criteria for Boolean Functions (I)",}
\href{http://www.ii.uib.no/~matthew/LCPartIf.pdf}
{http://www.ii.uib.no/\~{}matthew/LCPartIf.ps},
2004.

\bibitem{Rot:Bent}
O.S. Rothaus,
{"On Bent Functions",}
{\em J. Comb. Theory},
{\bf 20A}, pp. 300--305, 1976.

\bibitem{Rud:RS}
W. Rudin,
{"Some Theorems on Fourier Coefficients",}
{\em Proc. Amer. Math. Soc.},
No 10, pp. 855--859, 1959.

\end{thebibliography}
\end{document}